\documentclass[%
aip,
rsi,
amsmath,
amssymb,
reprint,%
]{revtex4-2}

\usepackage{amsmath,amssymb,amsfonts}
\usepackage{graphicx}
\usepackage{dcolumn}
\usepackage{bm}
\usepackage{braket}
\usepackage{siunitx}
\DeclareSIUnit{\mebibyte}{MiB}
\DeclareSIUnit{\kibibit}{Kib}
\DeclareSIUnit{\MSPS}{MSPS}
\DeclareSIUnit{\ppm}{ppm}
\DeclareSIUnit{\gauss}{G}
\DeclareSIUnit{\sqrtkh}{\ensuremath{\sqrt{\mathrm{kh}}}}
\usepackage[utf8]{inputenc}
\usepackage[T1]{fontenc}

\usepackage{mathptmx}
\usepackage{etoolbox}
\usepackage[hidelinks]{hyperref}
\usepackage{xcolor}

\makeatletter
\def\@email#1#2{%
 \endgroup
 \patchcmd{\titleblock@produce}
  {\frontmatter@RRAPformat}
  {\frontmatter@RRAPformat{\produce@RRAP{*#1\href{mailto:#2}{#2}}}\frontmatter@RRAPformat}
  {}{}
}%
\makeatother

\begin{document}
\title[]{A low-cost modular FPGA-based control system for neutral atom tweezer arrays}

\author{Zeyu Ye}
\affiliation{%
 Department of Physics, The University of Chicago, Chicago, Illinois 60637, USA}%
\affiliation{%
 Physics Division, Argonne National Laboratory, Lemont, Illinois 60423, USA}%
\author{Varun Jorapur}
\altaffiliation
 [Present Address: ]
 {IonQ, Inc., College Park, Maryland 20740, USA}
\affiliation{%
 Physics Division, Argonne National Laboratory, Lemont, Illinois 60423, USA}%
\author{Francesco Granato}
\altaffiliation
 [Present Address: ]
 {Department of Physics and Astronomy, Louisiana State University, Baton Rouge, Louisiana 70803, USA}
\affiliation{%
 Physics Division, Argonne National Laboratory, Lemont, Illinois 60423, USA}%
\author{Wanda Lindquist}
\altaffiliation%
  [Present Address: ]
  {Wyant College of Optical Sciences, University of Arizona, Tucson, Arizona 85721, USA}%
\affiliation{%
 Physics Division, Argonne National Laboratory, Lemont, Illinois 60423, USA}%
\author{David Peana}
\affiliation{%
 Physics Division, Argonne National Laboratory, Lemont, Illinois 60423, USA}%
\author{Isaac Both}
\altaffiliation%
[Present Address: ]
 {Department of Physics, Lake Forest College, Lake Forest, Illinois 60045, USA}%
 \affiliation{%
 Physics Division, Argonne National Laboratory, Lemont, Illinois 60423, USA}%
\author{Mouhamed Mbengue}
\altaffiliation%
[Present Address: ]
 {Department of Physics, The University of Chicago, Chicago, Illinois 60637, USA}%
 \affiliation{%
 Physics Division, Argonne National Laboratory, Lemont, Illinois 60423, USA}%
\author{Peter Mueller}
\affiliation{%
 Physics Division, Argonne National Laboratory, Lemont, Illinois 60423, USA}%
\author{Michael Bishof}
\affiliation{%
 Physics Division, Argonne National Laboratory, Lemont, Illinois 60423, USA}%
\affiliation{%
 James Franck Institute, The University of Chicago, Chicago, Illinois 60637, USA}%
 \email{bishof@anl.gov}

\date{\today}

\begin{abstract}

We present the Quantum Atom Control Kit (QuACK), a low-cost FPGA-based control system that provides memory-efficient precise signal timing and low-delay measurement-based decision making for operating neutral atom optical tweezer array experiments. The system consists of an FPGA module connected to a host computer via USB 3.0 and multiple cable-connected modular daughterboards enabling a myriad of configurable I/O formats. In this work, digital input and output as well as analog outputs are demonstrated. Multiple experimental sequence segments can be encoded with \qty{5}{\nano\second} timing resolution. On-chip condition kernels enable fast branching between segments based on digital inputs, eliminating host intervention for time-critical decisions, while high-speed USB communication allows rapid sequence updates and host-level orchestration. We demonstrate initial system loading and rearranging neutral ytterbium-171 atoms in optical tweezers. The combination of high-resolution timing, on-chip logic, and distributed orchestration architecture makes this system well suited for atomic, molecular, and optical (AMO) experiments pursuing applications in quantum information science.

\end{abstract}

\maketitle

\section{Introduction}
\label{sec:level1}

Modern atomic, molecular, and optical (AMO) physics experiments rely critically on precise and reconfigurable timing control. Typical experimental sequences span nanoseconds to minutes and involve tens to hundreds of control channels at multiple locations and in various formats, such as digital triggers, precision analog setpoints, and radio-frequency (RF) tones. Quantum applications require these experiments to scale to unprecedented levels of size and complexity. As such, timing systems need to evolve from simple deterministic playback of pre-compiled patterns to flexible architectures that support mid-sequence decisions \cite{reisenbauer2022}, spatially distributed devices, rapid reconfiguration for different tasks or optimization routines, and application-level asynchronous orchestration.

A common approach in AMO laboratories is to use a central digital pattern generator that replays hardware-timed waveforms on individual channels which either drive transistor-transistor logic (TTL) gates directly or trigger subsystems such as analog waveform generators, acquisition cards, or RF drivers \cite{keshet2013}. Existing hardware-timed control platforms include open-source systems such as ARTIQ \cite{artiq} and modular FPGA controllers \cite{bertoldi2020}, as well as commercial products such as OPX \cite{qmOPX}. These systems combine a hardware-timed core with interfaces for peripheral instruments.

In this work, we present the Quantum-Atom-Control-Kit (QuACK), a lightweight control system based on an Opal Kelly XEM7310-A200 FPGA module incorporating a Xilinx Artix-7 XC7A200T FPGA \cite{xem7310} with low-cost, modular daughterboards providing fast digital outputs, precision analog outputs, digital inputs and optional clock references. All daughterboards share common connector pinouts and are linked to the core via high-density microcoaxial cables, allowing flexible configurations with minimum hardware overhead compared to single board systems \cite{bertoldi2020} and sub-rack systems \cite{trenkwalder2021}, while still enabling high-speed communication with peripheral chips to generate output types other than digital pulses \cite{sitaram2021}.

For flexible sequence programming, our system allows multiple independent sequence segments stored in on-chip memory and accessed at runtime. Condition kernels on the FPGA enable on-chip decision-making and branching between segments based on digital inputs. High-speed USB 3.0 communication also allows low-delay interfacing with the host so that the system can be orchestrated with asynchronous peripherals and remote commands.

Our design also allows multiple FPGA modules to be combined to increase the total number of available channels or to act as local controllers at distant locations. The relatively low cost enables both scaling and the maintenance of independent test systems \cite{starkey2024}. At the application level, the system can be integrated into a distributed software framework within a local area network (LAN) via remote procedure calls (RPCs). This framework supports asynchronous configuration, logging, and high-level orchestration from Python, while the FPGA remains responsible for local deterministic timing. A graphical user interface (GUI) built in Python/PyQt6 \cite{pyqt} allows editing and visualization of sequences.

We demonstrate this system by operating a neutral atom experiment that arranges individual ytterbium atoms into defect-free optical tweezer arrays. The FPGA controls the full tweezer loading and detecting sequence. Host-level decision layer is used for atom-array rearrangement and optimization routines. The same architecture is being developed to perform analog quantum simulations and as a local quantum node controller in a quantum networking testbed, where on-chip decision-making is required to coordinate entanglement attempts.

The FPGA firmware, KiCad hardware design files, Python host software, and sequence visualizer for QuACK are provided in the Supplemental Material \cite{quack_repo}.

\begin{figure*}
\includegraphics[width=\linewidth]{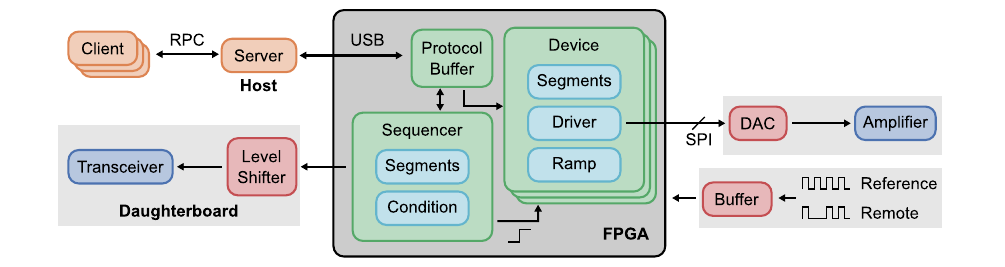}
\caption{\label{fig:hardware_connection} A simplified diagram of hardware connections around the FPGA module. It connects to a host computer through USB, which grants access to RPC commands. It also connects to various daughterboards through coaxial cables to interface with external devices.}
\end{figure*}

This manuscript is organized as follows:  Section II describes the hardware architecture, including the FPGA timing core as well as the designs and connections of daughterboards. Section III details the software design, including the action-based instruction format, condition kernels, and the networked device framework. In Section IV, we present example applications in neutral atom array preparation and feedback, and discuss timing performance and latency. We conclude in Section V with an outlook on extending this approach to larger modular systems and networked quantum experiments.

\section{Hardware architecture}
\label{sec:hardware}

Real-time control of AMO experiments requires that sequences of actions be executed with deterministic timing across a large number of channels. Microcontroller-based systems, though flexible in programming, do not scale easily to tens of parallel channels with nanosecond-level timing resolution \cite{starkey2024}. In contrast, FPGA-based systems provide parallelism, tight control over timing at the level of individual clock cycles, and sufficient logic resources to coordinate heterogeneous peripherals with low delay.

The FPGA module features a USB 3.0 interface for which the manufacturer reports measured transfer rates exceeding \qty{340}{\mebibyte\per\second} \cite{xem7310}. The module has also been used in real-time data acquisition systems \cite{delgadillo2022}. It has 124 I/Os and \qty{13140}{\kibibit} of block RAM (BRAM), which extends the number of available channels and depth of buffered sequences. On-board DDR3 memory can be used to further expand data storage. The FPGA is clocked at \qty{200}{\mega\hertz} by an on-board oscillator. For experiments requiring long coherence times, such as nuclear spin precession measurements \cite{bishof2016}, the timing machine can be disciplined to an external reference such as a rubidium frequency standard.

To interface with the experiment, the FPGA module functionality is expanded by a set of modular daughterboards. As a demonstration of this expansion, we implement digital outputs, digital counters, and precision analog outputs. All daughterboards connect to the FPGA module through 20-position \qty{50}{\ohm} microcoaxial cables (Samtec, FCF8-20-01-L), which offer low impedance and low skew for high-speed digital signaling. The daughterboards share a common connector pinout so they can be easily configured for different sets of I/O requirements. A representative hardware topology is shown in Fig.~\ref{fig:hardware_connection}.

The USB 3.0 link between the host computer and FPGA provides sufficient bandwidth for both streaming timing commands and transferring measurement data. In our implementation, this high-throughput link not only uploads timing sequences but also supports feedback without incurring significant dead time. This is particularly useful during rapid prototyping and for applications in which the FPGA must be tightly integrated with devices that natively communicate over USB, PCIe or Ethernet (e.g., scientific cameras or infrastructure orchestration). By design, we place the higher-level network functionality on the host computer rather than on the FPGA module itself. Similar capabilities have previously been realized directly on system-on-chip (SoC) platforms \cite{trenkwalder2021}, such as Xilinx Zynq devices, but delegating this layer to a general-purpose PC allows us to exploit mature software stacks and preserves FPGA resources for deterministic timing and parallel signal generation instead of protocol handling.

In the following subsections, we describe the main logic blocks implemented on the FPGA and the design of the digital and analog daughterboards.

\subsection{FPGA logic blocks}

\begin{figure}
\includegraphics[width=\linewidth]{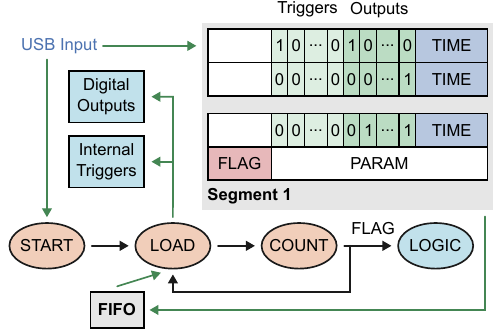}
\caption{\label{fig:timing_coding} Logic (black arrows) and data (green arrows) flows for sequence execution. Major stages of the main state machine are shown. The data in the active segment (Segment 1 for example) are loaded into a FIFO buffer. The state machine shifts them to the outputs and virtual triggers while it counts clock cycles for the target wait time encoded in ``TIME.'' End condition or branching logic is controlled by the FLAG and PARAM.}
\end{figure}

At the core of the timing system is a sequencer implemented in the FPGA. Experimental sequences are encoded as segments, each consisting of a table of actions. Each action is a time-state pair with a flag. The first element is a time interval, expressed as an integer number of FPGA clock cycles, and the second element is a representation of digital control over all channels. A simplified block diagram is shown in Fig.~\ref{fig:timing_coding}.  Digital channel outputs can be directly mapped from the sequencer table. Other channels implemented in serial protocols such as serial peripheral interface (SPI) use sequencer signals as triggers. Their drivers and data are wrapped in corresponding virtual device modules.

This representation is memory-efficient. Long idle periods are represented by single actions with large time intervals, while fast bursts of activity are represented by sequences of actions with small intervals. Because only state changes are stored, the required BRAM in the FPGA and the volume of data transferred from the host are significantly reduced compared to frame-based waveform representations.

Multiple segments can be stored in on-chip memory. In most cases, when the last action of a segment completes, the hardware retains the final output state and enters an idle state, awaiting either a new start command or a conditional trigger, until instructed otherwise by the host or a condition kernel. Indicated by the flag, condition kernels can be attached at designated decision points in a segment table. When the sequencer reaches such a point, the condition kernel evaluates its inputs and instructs the system to either move to the other segment or continue within the same segment. This enables branching logic and mid-sequence decisions with low latency, without host intervention.

Host-FPGA communication uses the Opal Kelly FrontPanel interface. It allows the host to upload and download segment tables, read status registers, and issue control commands including starting a segment by its index. Data transfer is pipelined through FIFO buffers. The USB 3.0 link allows the host to update hundreds of actions in a few milliseconds, including protocol overhead. This, in turn, makes shot-to-shot adaptation of segments feasible based on host-level analysis.

\subsection{Digital Outputs}

Digital outputs are used extensively in AMO experiments to control shutters, RF/microwave switches, camera triggers, arbitrary waveform generators (AWGs), and oscilloscopes. Many of these tasks require nanosecond-scale timing: for example, laser pulses used to excite atoms can have durations of a few tens of nanoseconds which must be shorter than the lifetime of the excited states \cite{li2025}, and their timing relative to other pulses must be controlled to similar precision for aligning to a detection window.

The Artix-7 FPGA supports general-purpose I/O rates in excess of \qty{200}{\mega\hertz}. In our design, digital outputs are grouped in banks of 16 channels on each daughterboard. Within the FPGA, outputs are updated synchronously by the sequencer on its main \qty{200}{\mega\hertz} clock. The nominal timing resolution is one FPGA clock cycle (\qty{5}{\nano\second}), and the output buffers need to have comparable speed and jitter performance.

For compatibility with the logic used for various daughterboards, the FPGA I/O banks used for extension are operated at \qty{3.3}{\volt} with LVCMOS33 standard. To interface with external devices that expect \qty{5}{\volt} TTL-level signals, each digital output is level shifted on the daughterboards. Level shifting and buffering are implemented using dual-supply bus transceivers (Texas Instruments, SN74LVC16T245) followed by \qty{25}{\ohm} bus transceivers (Texas Instruments, SN74BCT25245) that drive the coaxial outputs.

The absolute propagation delay in each channel is dominated by cable lengths and any intermediate connectors and gates. In practice, absolute delays can be compensated at the user end using calibrated cable lengths and temperature-stable coaxial cables if needed. The quantity of primary interest for most operations is the timing jitter of pulse intervals and standard deviation of channel-to-channel skew, rather than the absolute delay.

\begin{figure}
\includegraphics[width=\linewidth]{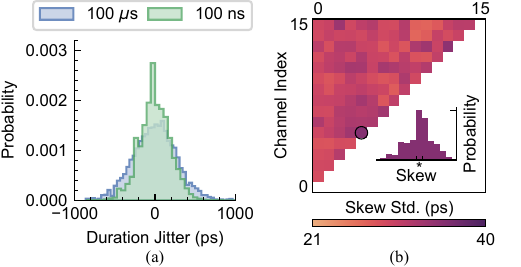}
\caption{\label{fig:digital_timing} Digital pulse output characterization. (a) Timing jitter of pulse durations for 100 $\mu$s and 100 ns pulses. (b) Standard deviation of skews between two channels on a daughterboard. Inset: histogram of skews in the worst channel-to-channel pair (circled) with bin size of \qty{15.8}{\pico\second}.}
\end{figure}

We characterized the jitter using an oscilloscope (Tektronix, DPO 5204) with \qty{50}{\ohm} BNC patch cables connected to a digital daughterboard. Our output stages provide typical rise time of \qty{4.97 \pm 0.12}{\nano\second} and fall time of \qty{2.645 \pm 0.041}{\nano\second}. For each channel, we generated sequences of \qty{100}{\nano\second} and \qty{100}{\micro\second} wide pulses and measured the variations in pulse duration. The standard deviation of the pulse width was found to be below \qty{1}{\nano\second} across all channels, as shown in Fig.~\ref{fig:digital_timing}(a). We also measured the skews between pulses on different channels. The standard deviation in channel-to-channel skews are all also below \qty{1}{\nano\second}, as shown in Fig.~\ref{fig:digital_timing}(b), demonstrating that the daughterboard and FPGA logic preserve sub-nanosecond relative timing across output channels.

For applications requiring true rail-to-rail \qtyrange{0}{5}{\volt} swings and tighter control of output impedance, we have also implemented an alternative variant in which the outputs are buffered by high-output-current rail-to-rail operational amplifiers (Texas Instruments, OPA4354). In that configuration, the rise time is \qty{19.54 \pm 0.47}{\nano\second} and fall time is \qty{20.54 \pm 0.49}{\nano\second}, so it is used primarily for slower digital gating and switching rather than for the shortest pulses.

\subsection{Digital Inputs}

Digital inputs often form the backbone of experiments that require synchronized clocking, triggering an on-board logical sequence to an external event, pulse counting, or fast inter-device communication. Due to the simplicity of binary signal processing compared to analog-to-digital conversion, digital inputs are also optimally adapted to nanosecond-scale response times.

Daughterboards with 8 digital input channels were implemented. Incoming signals are buffered and level-shifted through Texas Instruments SN74LVC8T245 ICs with optional termination resistors. In applications as counters, corresponding FPGA pins are configured as inputs for \qty{3.3}{\volt} pulses, which count the numbers of pulses detected within internal-gate-controlled time ranges. While theoretically (given the \qty{5}{\nano\second} clock cycle), this digital input scheme should be able to count up to around \qty{100}{\mega\hertz}, we have only tested it up to \qty{50}{\mega\hertz} (up to which clock-jitter-limited performance is maintained). At 50 MHz, the measured counts differed from the nominal expectation of \num{e+9} counts by at most 7 counts across \qty{20}{s} acquisitions, corresponding to a maximum absolute fractional deviation of \num{7e-9}.

Combined with the logic block implemented with the sequence end conditions, those inputs can be used for fast branching based on remote pulses without any hardware modifications, and its low jitter is beneficial for cross-device synchronization.

\subsection{Analog Outputs}

\begin{figure}
\includegraphics[width=\linewidth]{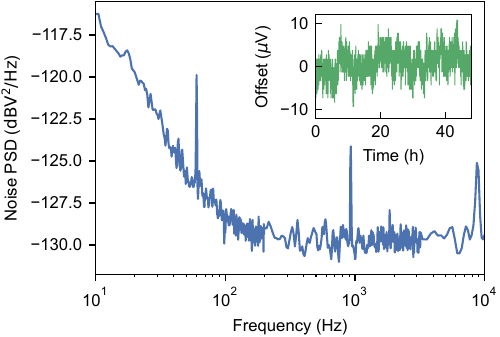}
\caption{\label{fig:dac_noise} DAC output noise performance. Transient noise spectrum is shown in terms of voltage noise power spectral density. Inset: Long-term setpoint drift over \qty{48}{\hour}.}
\end{figure}

Precision analog outputs with long-term stability are commonly required as setpoints for controlling applied electromagnetic fields.  Some common examples include laser intensity stabilization, current stabilization for magnetic field coils, and electrode voltage control.  
For example, slow drifts in the magnetic field can lead to dephasing in Raman pulse operations on nuclear spins. 

The analog output daughterboards provide up to five independent channels of analog outputs per daughterboard, with a maximum update rate of \qty{1}{\MSPS} per channel. Each channel is based on a low-temperature-drift \qty{0.05}{\ppm\per\degreeCelsius}, serial-input digital-to-analog converter (DAC, Analog Devices, AD5542A). The DACs use a \qty{2.5}{\volt} precision voltage reference (Analog Devices, LTC6655-2.5, MSOP-8). For this package, the datasheet specifies a typical long-term drift of \qty[per-mode=symbol]{60}{\ppm\per\sqrtkh}. The maximum temperature coefficient is \qty{2}{\ppm\per\degreeCelsius} \cite{ltc6655}. The reference is buffered using a low-noise linear regulator (Analog Devices, LT3042) to support multiple DACs. The DAC outputs are then amplified and buffered with high precision operational amplifiers (Analog Devices, ADA4510-2), which have $\pm$\qty{5}{\micro\volt} typical offset and $\pm$\qty{70}{\nano\volt\per\degreeCelsius} typical offset drift. In our setup, outputs are amplified to $\pm$\qty{10}{\volt}. To maintain low noise on the analog outputs, all DACs and analog buffers on the daughterboard are powered by ultra-low-noise low-dropout regulators (LDO, Analog Devices, LT3045 and LT3094), which provide high power-supply rejection over a broad bandwidth (\qty{<= 1}{\mega\hertz}), covering the bandwidth of most in-lab servo loops. Figure~\ref{fig:dac_noise} shows a representative output noise spectrum of a DAC channel measured with a spectrum analyzer (SRS, SR780) and its long-term drift around its mid-scale output (about \qty{0}{\volt}), recorded over \qty{48}{\hour} with a digital multimeter (Fluke 8842A) with board ambient temperature swinging within \qtyrange{20.1}{22.4}{\degreeCelsius}.

The analog daughterboards share the same connector and cabling as the digital daughterboards. In the FPGA, the corresponding digital I/O pins are reconfigured as multiple parallel SPI buses. A typical five-channel daughterboard uses 15 FPGA pins to support five separate SPI buses.

To integrate the DAC in the timing sequence, the encoded SPI commands in a segment are pre-defined in a RAM block which is wrapped in a module for each channel. This module acts like a virtual device inside the FPGA and is triggered by the digital tables. Each time it is triggered, it transfers the next SPI command through the SPI driver module inside. To allow direct access from the host for initialization or tests, the SPI driver is multiplexed so that the host can directly transfer commands outside of a sequence for debugging.

In many applications, analog outputs are changed slowly, for example during adiabatic ramps of trap depths. To avoid storing large numbers of intermediate values in memory, each DAC channel can utilize an on-board ramp processing core. It interpolates between successive values at a fixed \qty{1}{\mega\hertz} update rate using a linear increment.

This serial coordinator structure can be extended to designs beyond this DAC chip by replacing the implementation inside of its driver module. For example, RF daughterboards for AOMs can be built with Direct Digital Synthesis (DDS) chips such as the Analog Devices AD9910.

\section{Software}

Many experiments require rapid reconfiguration between different operating modes, such as calibration, debugging, and data acquisition. Rather than relying exclusively on pre-loaded sequences stored in FPGA memory, our design supports both on-board storage of reusable segments and dynamic upload of segments from the host during runtime. For typical experiments, the USB 3.0 link allows transfer of updates to segments or parameters without adding significant dead time.

The hardware described in Sec.~\ref{sec:hardware} provides deterministic timing and low-level control of digital and analog channels. To make this capability broadly usable in a complex apparatus, we built a flexible software stack on top of the FPGA, which consists of a host-side control server, a networked device framework that exposes the FPGA and other hardware as remote services, and graphical tools for editing and visualizing sequences.

\subsection{Host communication and sequence management}

On the host, a Python/C++ server using the Opal Kelly FrontPanel API is responsible for communicating with the FPGA module. This server provides a set of high-level commands for uploading segment tables, reading status or data registers, and starting or stopping the timing engine. When low latency is not critical, a Python server can be used for convenience. The server can also be implemented as a C++ library to be integrated into a more complicated application such as a local service with atom array rearrangement.

Sequences are assembled on the host from human-readable descriptions of actions. Each of them is defined as a dataframe with each row as a time step and each column as a channel. The first column is reserved for the time durations of the steps, and the first row is reserved for channel names, and nicknames can be noted as the column names. The output behavior is described by cell values in the format corresponding to the types of the channels. For example, within the column of an analog channel, each step has a float value in unit of volts. For float values, a special ">" marker appended to a value indicates the start of a ramp toward the value in the next column. If there is no new value in the row with the ramp marker, it continues the on-going ramp. The server resolves the time intervals, compiles the dataframe into a sequencer table and, for the analog channels, into appropriate SPI commands as well. Compiled segment tables are then uploaded to the FPGA and stored in block RAM. Because each action can specify updates on multiple channels, the action representation tends to remain compact when many channels are used. Those sequences can be later called by their indexes from the server.

\subsection{Remote procedure calls}

Usually, the FPGA timing system is only one component among many devices that must be monitored and controlled, including laser controllers, stabilization servos, environmental sensors, and safety interlocks. Managing this heterogeneous set of devices with a uniform interface is challenging because of the variety of data formats, communication protocols, and interdependencies. We therefore adopt a distributed, network-based approach in which each device, or virtual device, runs as an independent process accessible via RPCs.

Most devices are wrapped in Python servers that use RPyC over the LAN \cite{rpyc}. RPyC does not require a separate interface-definition or code-generation step and allows remote services and methods to be accessed directly through Python names and attributes. This reduces auxiliary files, making the system easier to modify and deploy during rapid development for laboratory control. The server wraps the low-level APIs into uniform parameter objects. The device servers are contained in individual processes within a device manager, which loads/unloads devices in the same Python environment. The managers can be deployed on different machines, including single-board computers such as Raspberry Pi, and can be running with different dependencies.

A registry server acts as a central directory. When a device connects, it registers itself with the registry, allowing clients to discover it later, and they can set up direct connections to the device servers. This decouples the experiment script from the specific network locations. Monitoring data from devices are logged in InfluxDB, a time-series database \cite{influxdb}. This enables continuous monitoring and retrospective analysis of system parameters such as laser powers and temperatures. Other helper servers can be implemented as virtual devices under this architecture. For example, a lightweight monitor process which subscribes to the database or specific devices directly and sets global error flags, which can be checked by the experiment scripts to pause the experiment if needed. An additional layer based on gRPC \cite{grpc} can be used to provide cross-language interfaces, for example between C++ and Python, or interface with other experiments. A set of devices can be wrapped into an experiment core, which can be accessed from an experiment script, a local optimizer, or remote commands.

\subsection{User interface}

\begin{figure}
\includegraphics[width=\linewidth]{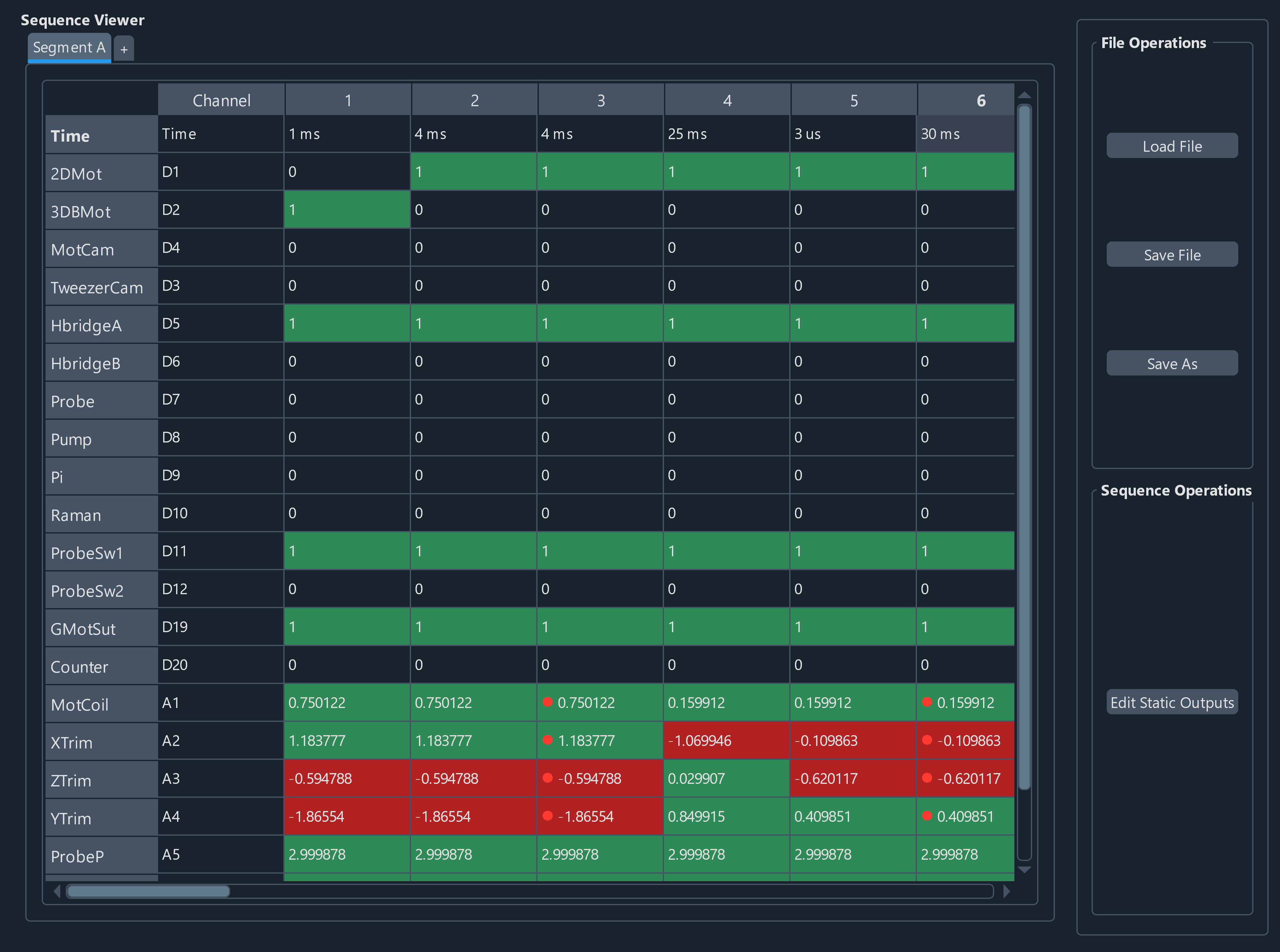}
\caption{\label{fig:gui_screenshot} Example of a sequence file loaded into the software GUI. Starts of ramps are marked with red dots. The special marks for ramps can be set/unset with the right-click menu.}
\end{figure}

While most experiment control and optimization tasks are scripted, a graphical user interface (GUI) is valuable for constructing, inspecting, and modifying sequences, particularly during initial development and debugging. Sequence files are usually stored in JSON format, which is both human-readable and easily parsed. We have developed a simple cross-platform GUI editor in Python using PyQt6. The GUI displays sequences in a transposed two-dimensional table with an example shown in Fig.~\ref{fig:gui_screenshot}. In practice, we often use the GUI to draft or visualize sequences and then update them in scripted form, or to inspect sequences generated by optimization routines.

\section{Example applications}

To demonstrate the capabilities of the control system, we present two representative applications: (i) neutral atom array loading and rearrangement in an ytterbium experiment and (ii) a brief discussion of its role as a quantum network orchestrator. The first illustrates deterministic timing combined with host-level feedback, while the second highlights the potential for coordinating entanglement attempts across distributed nodes.

\subsection{Atom array loading and rearrangement}

In our setup, neutral ${}^{171}\mathrm{Yb}$ atoms are prepared in optical tweezers and their ground nuclear spin states are probed and manipulated by fast laser pulses. Required control signals of the loading and operation sequences are generated by the FPGA system. Communication between the hardware and the host is used to perform feedback on atom arrays based on readouts of atoms.

\begin{figure}
\includegraphics[width=\linewidth]{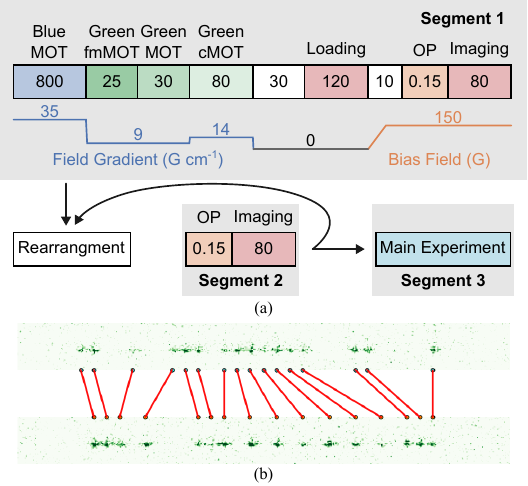}
\caption{\label{fig:experiment_sequence} Test sequence for loading and rearranging Yb atom arrays controlled by the device. Durations are given in \unit{\milli\second}. (a) Multiple sequence segments for random access are shown. In Segment 1, the magnetic field generated by coils in vertical direction is shown, which is controlled by 1 analog channel as coil current reference and 2 digital channels for H-bridge control. (b) One-shot images of atoms before and after the rearrangement.}
\end{figure}

Ytterbium atoms from a homemade oven at \qty{\approx 350}{\degreeCelsius} are cooled using the ${^{1}S_{0}} \leftrightarrow {^{1}P_{1}}$ transition at \qty{399}{\nano\metre} ($\Gamma/2\pi$ = \qty{29}{\mega\hertz}) through transverse cooling, Zeeman slowing, and a \ang{30} bending 2D magneto-optical trap (MOT). Then, they are captured by a blue 3D MOT in an ultra-high vacuum (UHV) glass cell. MOT beams have $0.28I_{\text{sat}}$ beam intensity and $-0.5\Gamma$ detuning. The \qty{399}{\nano\metre} transition has a Doppler temperature of \qty{699}{\micro\kelvin}. To load atoms into \qty{\approx 800}{\micro\kelvin} tweezers, atoms are transferred to green MOTs using the narrow ${^{1}S_{0}} \leftrightarrow {^{3}P_{1}}$ transition at \qty{556}{\nano\metre} ($\Gamma/2\pi$ = \qty{182}{\kilo\hertz}). To increase the efficiency of this handoff, atoms are initially captured in a broadened green MOT with the magnetic field gradient dropped from \qty{35}{\gauss\per\centi\metre} to \qty{9}{\gauss\per\centi\metre}. The MOT is broadened by both intensity ($114I_{\text{sat}}$) and frequency which is modulated between $-21.5\Gamma$ and $-5.6\Gamma$ using digital frequency ramps in a DDS module. Then, atoms are transferred into a single frequency green MOT with $32I_{\text{sat}}$ beam intensity and $-5.2\Gamma$ detuning and cooled for \qty{30}{\milli\second}. For optimal MOT loading, a compression stage is performed. The magnetic field gradient is ramped up to \qty{14}{\gauss\per\centi\metre}, atoms are further cooled with $4I_{\text{sat}}$ beam intensity and $-5.2\Gamma$ detuning at the location overlapped with the tweezers. There are three pairs of trim coils used for adjusting the positions of the MOTs. The stage durations, intensities, detunings and trim fields were optimized using machine learning based on Gaussian process regression (GPR). The GPR optimizer is running as an RPC client of an experiment server which connects to the equipment. Optimization of experiment sequences is performed regularly to keep their robustness.

\begin{figure}
\includegraphics[width=\linewidth]{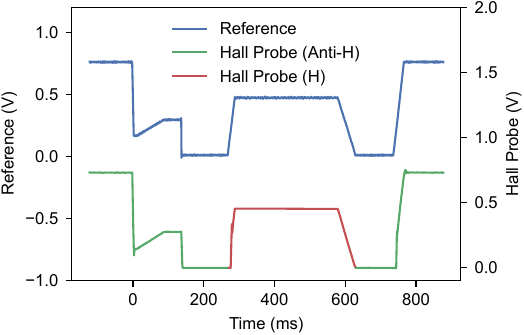}
\caption{\label{fig:magnetic_field_waveform} Magnetic field waveform reference signal generated by the controller and the measured current on the coils. Anti-H stands for anti-Helmholtz, and H means Helmholtz. An H-bridge is used to switch between the two configurations, and, in both configurations, coil current is regulated to the reference with a fast servo.}
\end{figure}

The loading sequence is stored in one segment of the FPGA sequence memory, shown in Fig.~\ref{fig:experiment_sequence}. Digital channels are used to control the shutters and H-bridges for laser and current switching. Analog waveforms are generated for referencing the MOT coil currents (shown in Fig.~\ref{fig:magnetic_field_waveform}), the trim coil currents, and the green MOT laser intensities. The DDS module controls RF frequencies sent to a double-passed acousto-optic modulator (AOM) which sets the MOT detuning. To show the functionality of this control, the Hall probe (Danisense, DS200ID) reading of the MOT coils is shown in Fig.~\ref{fig:magnetic_field_waveform}. The coil current is sensed by a high precision foil resistor (Vishay, VCS1625ZP) on the Hall probe output and is stabilized by a homemade servo based on a signal processing board (Red Pitaya, STEM 125-14).

Optical tweezers operate at \qty{\approx 759}{\nano\metre}, close to the "magic" wavelength of the narrow ${^{1}S_{0}} \leftrightarrow {^{3}P_{0}}$ clock transition \cite{lemke2009}. Individual trapping beams are generated by a 2-axis acousto-optic deflector (AOD, AA Opto-electronic, DTSXY-400-760) acting on the output of a tapered amplifier (TA) after spatial filtering through an optical fiber. Tightly-focused optical tweezers are created with a NA=0.63 custom objective (Special Optics). Dynamic control of the waveform driving the AOD is implemented on a PCIe AWG (Spectrum Instrumentation, M4i6622-x8), which is connected to the same computer as the FPGA module.

After the MOT is extinguished, a \qty{120}{\milli\second} loading pulse sharing the same path as the MOT beams is used for cooling and light-assisted collisions with $1.6I_{\text{sat}}$ and $-1\Gamma$ detuned from $\ket{{^{1}S_{0}},m_{F}=\pm 1/2} \leftrightarrow \ket{{^{3}P_{1}},F'=3/2,m_{F'}=\pm 3/2}$. The magnetic field is then ramped up to \qty{150}{\gauss}. Atoms are imaged by exciting the $\ket{{^{1}S_{0}},m_{F}=-1/2} \leftrightarrow \ket{{^{3}P_{1}},F'=3/2,m_{F'}=-3/2}$ transition, and optical pumping into 
$\ket{{^{1}S_{0}},m_{F}=-1/2}$
through $\ket{{^{1}S_{0}},m_{F}=+1/2} \leftrightarrow \ket{{^{3}P_{1}},F'=3/2,m_{F'}=-1/2}$ is performed before each image. The filling fraction is \num{\approx 0.7} and tweezer lifetime is measured to be \qty{40 \pm 5}{\second}. In the application of quantum networking, to maximize the attempt rate, defect-free arrays of requested patterns are desired \cite{li2025}. Such arrays can be achieved by programming waveforms to move frequency tones of corresponding atom traps and rearranging atoms in stochastically loaded arrays \cite{endres2016}.  Atom fluorescence is collected with a scientific CMOS (sCMOS) camera (Teledyne, 01-PRIME-BSI-EXP). With \qty{80}{\milli\second} exposure time, atom occupancy can be distinguished with fidelity \qty{>99}{\percent}. For each feedback loop for atom rearrangement, a camera trigger is issued from the FPGA, which is synchronized with the RF switch and intensity stabilization reference of the imaging beam. Then, the computer reads the frame from the camera through USB and calculates the tweezer occupancy. Then, it generates the desired waveform and plays it on the AWG. Once it is done, the computer starts another imaging segment on the FPGA and checks the result again from the camera. It can then decide the next segment to be executed.

We tested atom rearrangement from 30-site arrays to 10 site defect-free arrays with \qty{\approx 2.9}{\micro\metre} spacing. Infidelity can be caused by beating between close frequency tones during movement and atom loss due to ${^{3}P_{1}} \rightarrow {^{3}S_{1}}$ off-resonance scattering during imaging \cite{lis2023}. We believe that atom survival can be improved with a shorter exposure time, or with a different tweezer wavelength for imaging \cite{ma2023}. For rearranging to a fixed size of subarrays, extra atoms can be stored in a reservoir zone and pulled from the reservoir to fill the defects in the first run to further improve the experimental duty cycle. Such decisions can be made by the computer in our design to determine which segment will be played next. The typical delay for a computer decision is \qty{275 \pm 19}{\micro\second}. In future experiments, a continuously loaded reservoir can be used to maintain the experiment arrays \cite{chiu2025}, and the computer can orchestrate its maintenance between tasks. 

\subsection{Quantum network orchestrator}

Beyond local control of a single apparatus, quantum networking experiments require coordination between multiple spatially separated nodes, often connected over optical fibers, requiring synchronization between fast entanglement attempts and coordination using slower classical communication links. In this context, the control hardware must function as a real-time, node-level orchestrator that can trigger local operations with precise timing, react rapidly to detection events, and follow schedules dictated by a higher-level network protocol \cite{chung2026}.

Our FPGA control system can be applied as a prototype quantum network node controller. At each node, the FPGA executes local control sequences for entanglement generation attempts, including preparation of a local qubit and emission of photons into a fiber. Condition kernels monitor digital inputs connected to single-photon detector outputs and, upon detection of an event, branch to segments implementing subsequent local operations such as storage, retrieval, or local gate sequences.

At the same time, the host-level software communicates with a centralized orchestrator of a local experiment or peer nodes using RPCs over the LAN to execute distributed quantum networking experiments. Because the FPGA timing engine can be commanded at the segment level and can perform mid-sequence branching, it is well suited to this division of labor: the FPGA handles deterministic sequence timing and local decisions, while the network protocol handles the application layer with instructions such as which pairs of nodes should attempt entanglement \cite{yu2025}.

Although a full experimental demonstration of multi-node entanglement is beyond the scope of the present work, we have verified in test configurations that two FPGA modules at different locations can be synchronized via digital inputs and that trigger and response latencies are compatible with the time scales of heralded entanglement protocols based on atomic emission of optical photons \cite{li2025}.

\section{Conclusions and outlook}

We have developed QuACK, an FPGA-based control system that provides precise, flexible experimental control for AMO and quantum networking experiments through an action-based timing architecture with on-chip decision-making capabilities. Built around a commercial FPGA module with USB 3.0 connectivity, the system is extended by modular daughterboards providing fast digital outputs and inputs, as well as precision analog outputs.

The action-based timing engine encodes sequences as time intervals paired with output updates, significantly reducing memory usage compared to frame-based representations while maintaining \qty{5}{\nano\second} timing resolution. Multiple sequence segments can be pre-loaded into on-board memory, with condition kernels enabling sub-microsecond branching between segments based on digital inputs without host intervention for time-critical decisions. The modular hardware design allows flexible channel configuration through interchangeable daughterboards connected via high-speed microcoaxial cables, achieving sub-nanosecond jitter for digital outputs and providing low-drift analog control essential for magnetic field and laser power stabilization.

The system integrates seamlessly into a distributed software framework where devices are exposed as network services via RPCs, enabling uniform control interfaces and real-time monitoring. This architecture separates deterministic timing from higher-level orchestration, allowing experiment scripts and machine-learning optimizers to interact efficiently with the hardware while the FPGA maintains precise timing control.

We demonstrated these capabilities in a neutral atom experiment with ${}^{171}\mathrm{Yb}$ arrays in optical tweezers, where the system controls complete atom loading and rearrangement. The combination of host-level feedback and on-chip decisions enables real-time atom-array rearrangement, optimization of loading parameters, and fast branching based on external conditions. The communication latency is negligible compared to camera exposure and computation times.

The architecture readily extends to additional applications: more daughterboards can expand channel counts and synchronization through digital inputs enables distributed nodes for quantum networking. The same combination of action-based timing, on-chip logic, and networked control could prove valuable beyond AMO physics in any application requiring precise timing, flexible feedback, and integration with heterogeneous instrumentation.

\section*{Data Availability}
The data that support the findings of this study are available from the corresponding author upon reasonable request.

\section*{Acknowledgment}
The authors thank Joaquin Chung a careful reading of the manuscript and for providing helpful feedback.

\appendix

\bibliography{main}

\end{document}